\documentclass[onecolumn,10pt]{article}
\usepackage[top=.75in, bottom=.75in, left=.75in, right=.75in]{geometry}
\usepackage{amsmath,amsfonts,amscd,amssymb}
\usepackage{graphicx}
\usepackage{epstopdf}
\usepackage{overpic}
\usepackage{cancel}
\usepackage{rotating}
\usepackage{url}
\usepackage{caption}
\usepackage{color}
\usepackage{rotating}
\usepackage{multirow}
\usepackage{wrapfig}
\usepackage{mathtools}
\usepackage{subeqnarray}
\usepackage{setspace}
\usepackage{palatino} 
\usepackage[numbers,sort&compress]{natbib}

\usepackage[bottom,flushmargin,hang,multiple]{footmisc}
\usepackage{lipsum}
\newcommand\blfootnote[1]{%
  \begingroup
  \renewcommand\thefootnote{}\footnote{#1}%
  \addtocounter{footnote}{-1}%
  \endgroup
}

\definecolor{header1}{cmyk}{0,0,0,1}

\usepackage[utf8]{inputenc}

\usepackage[normalem]{ulem}
\usepackage{color}

\title{\vspace{-.45in}{\huge\selectfont \textbf{The transformative potential of machine learning for experiments in fluid mechanics}}\vspace{-.15in}}

\author{\normalsize{Ricardo Vinuesa$^{1,2*}$, Steven L. Brunton$^{3}$ and Beverley J. McKeon$^{4}$}\\
\footnotesize{$^1$FLOW, Engineering Mechanics, KTH Royal Institute of Technology, Stockholm, Sweden}\\
\footnotesize{$^2$Swedish e-Science Research Centre (SeRC), Stockholm, Sweden}\\
\footnotesize{$^3$Department of Mechanical Engineering, University of Washington, Seattle, WA, USA}\\
\footnotesize{$^4$Department of Mechanical Engineering, Stanford University, Stanford, CA 94305
\vspace{-.2in}}
}

\date{}

\begin{document}
\maketitle

\blfootnote{$^*$ Corresponding author: rvinuesa@mech.kth.se}
\vspace{-.2in}
\begin{abstract}
The field of machine learning has rapidly advanced the state of the art in many fields of science and engineering, including experimental fluid dynamics, which is one of the original big-data disciplines. This perspective will highlight several aspects of experimental fluid mechanics that stand to benefit from progress advances in machine learning, including: 1) augmenting the fidelity and quality of measurement techniques, 2) improving experimental design and surrogate digital-twin models and 3) enabling real-time estimation and control.  
In each case, we discuss recent success stories and ongoing challenges, along with caveats and limitations, and outline the potential for new avenues of ML-augmented and ML-enabled experimental fluid mechanics. 
    

\end{abstract}


\section*{Introduction}\label{sec:intro}


The current renaissance in machine learning has the potential to revolutionize many aspects of the human experience, including scientific discovery.  The ability to learn from data is developing synergetically with the generation of big data in a range of applications, as well as rapid advances in materials science and additive manufacturing.  In this article, we describe some frontiers impacted or opened by machine learning (ML) applied to experimentation in fluid mechanics, a discipline at the core of many applications and recent technological developments in health, transportation, energy, and the environment. 

From da Vinci’s sketches of eddies forming in a pool to present-day qualitative and quantitative observations of flow through or over complex geometries, experiments are often either the first means of discovery, or the only means to reach extreme flow conditions in advance of numerical development. They may serve to complement, validate or extend theory and simulation. However, the benefits of observing the ``truth'', i.e. the real-world manifestation of flow phenomena, are usually accompanied by challenges associated with non-ideal experimental conditions (noise, vibration, thermal drift, etc.) and/or diagnostic limitations (limited spatial and/or temporal resolution, sensor accuracy, bias, etc.). For example, the study of turbulence near walls, of fundamental importance to a range of engineering applications as well as a foundational physics problem, is hampered by the physical dimensions of even the smallest (intrusive or non-intrusive) probe as the Reynolds number increases and the smallest turbulent scales decrease in size.  This particular problem has given rise to a range of empirical correlations, which may then be used in efforts to differentiate nuanced scaling arguments.
In industrial settings, experiments form critical design and testing steps, even in the age of advanced computational capabilities. ML has been increasingly adopted as a core technology in the field~\cite{Taira2017aiaa,Brunton2020arfm,Brenner2019prf}, with a focus on computational fluid dynamics (CFD)~\cite{vinuesa_brunton}. In this study we consider the potential impact of ML on experimental capabilities with a particular emphasis on the multiscale nature of turbulent flow, which may pose specific challenges to classical approaches.

We divide the experimental-fluid-mechanics applications into three different categories, ordered in terms of increasing novelty of the ML-enabled capability to the field: 1) augmentation of the fidelity of measurement techniques, for example by replacing empirical corrections with generalizable schemes learned from data and increasing what can be learned from observation through super-resolution or noise removal; 2) enhanced modeling, active learning, and experimental design, including digital twins; and 3) capabilities potentially enabled by AI/ML such as estimation and control. For all three categories we identify various machine-learning (ML) methods which can improve the experimental method. 

\begin{figure}
\centering 
\includegraphics[width=0.995\textwidth]{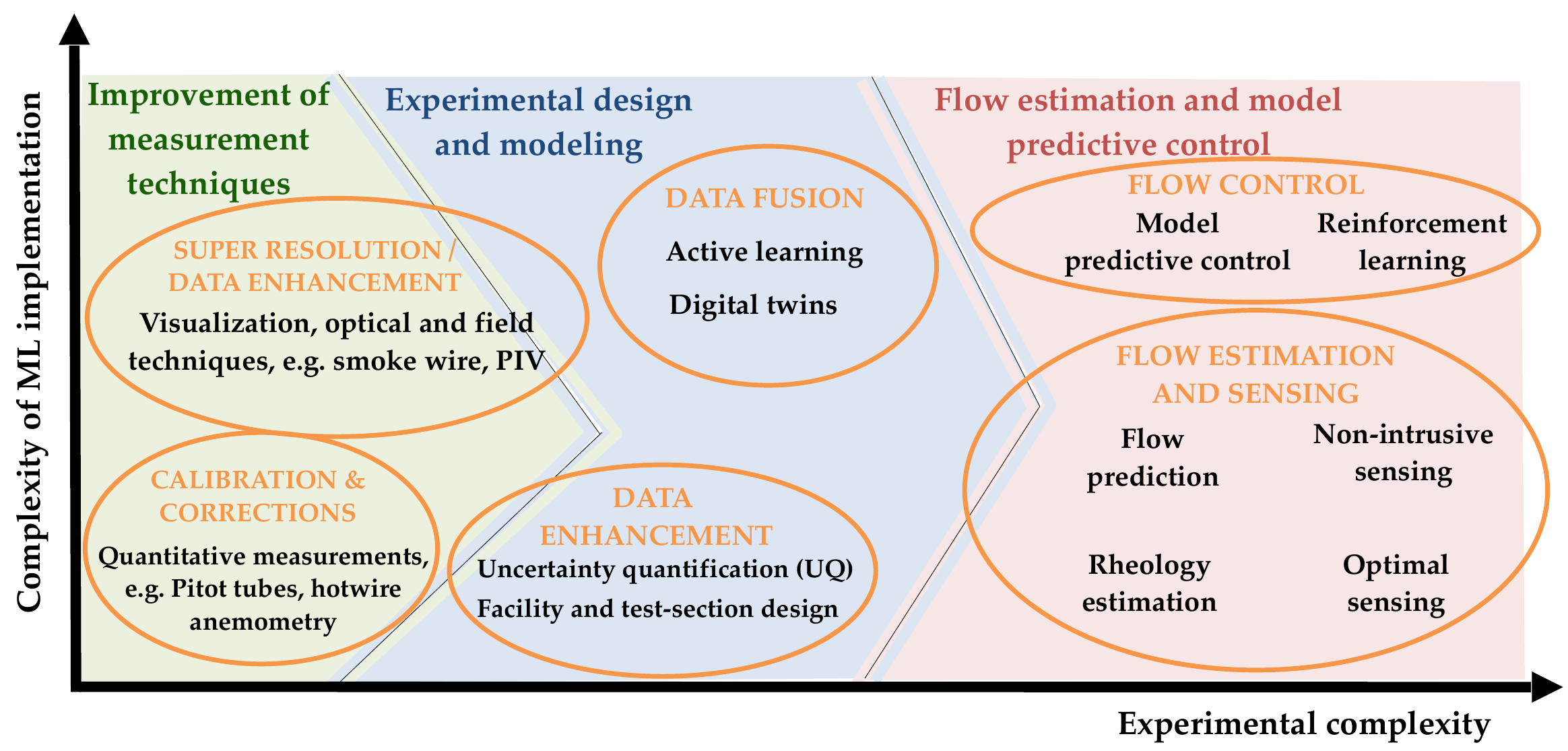}
   \caption{{\bf Summary of ML impact on experimental fluid mechanics.} Classification of all the experimental applications in terms of their main category, namely (green) measurement techniques, (blue) experimental design and modeling, as well as (red) estimation and control. Inside each region we show in black experimental techniques or tasks, whereas the orange bubbles indicate general improvements brought to those tasks by ML. The boundaries between categories are designed as arrows, reflecting the possible flows among them. All the examples are ranked by their ML and experimental complexities, as discussed in more detail throughout the manuscript. Note that this assessment is not exhaustive and is meant to illustrate possible applications.}
   \label{fig:summary}
\end{figure}

A summary of the analyzed experimental methods for each of the categories is presented in Figure~\ref{fig:summary}, together with potential areas where ML can help. We also include a classification of method difficulty, both in terms of experimental complexity and the challenges associated with the coupled implementation of the ML method and the experimental technique. Note that this classification is illustrative, and it is associated with the examples provided in the text; it is therefore not meant to be a categorical statement regarding the merits of any of the experimental or data-driven methods. Furthermore, the considered data points are taken as a combination of an ML method applied to a particular experimental technique. For example, if convolutional neural networks (CNNs)~\cite{lecun} are used for non-intrusive sensing, the complexity of this ML method is lower than if CNNs are used for sensing and control; even if the method is the same, its implementation in a more challenging application requires a more cumbersome algorithmic solution. This will be discussed in further detail throughout the manuscript.






\section*{Improvement of measurement techniques by means of machine learning}

Machine learning  offers the potential to improve existing measurement techniques in a range of ways, from learning precise forms of corrections required due to sensor imperfections and bias, to data imputation, i.e. providing insight into data that was not measured, for example due to limited sensor resolution.  We begin by describing these examples and the ML approaches that have been used to augment classical techniques.

One important problem of the experimental studies in fluid mechanics is the error introduced by the instrumentation employed to perform the measurements. ML has the potential to learn the underlying transfer function between the variable under consideration and the observed value, even in measurements with bias due to practical sensor limitations or noise. 

One widely-used sensor to measure velocity profiles is the so-called Pitot tube, in which the difference between total and static pressures is used to obtain the flow velocity through the classical Bernoulli equation. These sensors do not have a sufficiently fast frequency response for measuring the turbulence fluctuations, but they can provide robust measurements of the mean flow~\cite{bailey_et_al}. However, these sensors require a number of corrections~\cite{tavoularis}, where McKeon {\it et al.}~\cite{mckeon_et_al} proposed corrective terms for viscous effects, shear, wall interference and turbulence. While there is a physical motivation behind these corrections, their functional forms are empirical~\cite{vinuesa_corrections}, a fact that hinders their usage for general purposes. Similar problems are identified when correcting the position of the probe with respect to the wall~\cite{orlu_wall,vinuesa_wall} and when it comes to accounting for the filtering effect on the fluctuations measured by a hot-wire anemometer~\cite{ashok_hw}, where empirical approaches are employed~\cite{hw_melbourne,hw_iit}. In the latter case, an important open question which limits the applicability of the correction schemes to various flow cases revolves around the lack of universality of the small scales in flows with pressure gradients. In this sense, being able to exploit existing databases and novel machine-learning methods to obtain more sophisticated correction strategies has the potential to improve experimental measurements. Interpretable correction models may be learned using symbolic-regression strategies such as genetic programming~\cite{gep} or the sparse identification of nonlinear dynamical systems (SINDy)~\cite{Brunton2016pnas}. Alternative approaches, such as the DeepONet framework~\cite{lu2021learning}, make it possible to learn nonlinear correction operators from data. In this sense, DeepOnet could be used to generate corrected mean and fluctuating velocity profiles as a function of a number of input parameters, based on the boundary-layer development and integral quantities.

Non-quantitative measurement techniques provide visual information regarding the state of the flow, but no numerical information regarding its quantities of interest. An example of this type of measurement is the so-called smoke-wire visualization technique~\cite{smoke_wire}, which enables obtaining insight into a number of flow characteristics, such as the transition or separation locations. One possibility to augment the measurements from smoke-wire visualizations is to use numerical data of the same flow case to train computer-vision tools, {\it e.g.} convolutional neural networks (CNNs)~\cite{lecun}. With these databases, it would be possible to establish a mapping between the visualization (which does not include any quantitative data) and the velocity and pressure vectors obtained from the simulation. Due to the strong spatial correlations present in turbulent fluid flows, CNNs can be used to effectively predict patterns~\cite{guastoni2}, being able to add the quantitative information to the flow visualizations. In fact, other deep-learning approaches enable to first enhance the resolution of the visualization~\cite{guemes_gans,super1,super2}, and subsequently add the quantitative information.

\begin{figure}
\centering 
\includegraphics[width=0.8\textwidth]{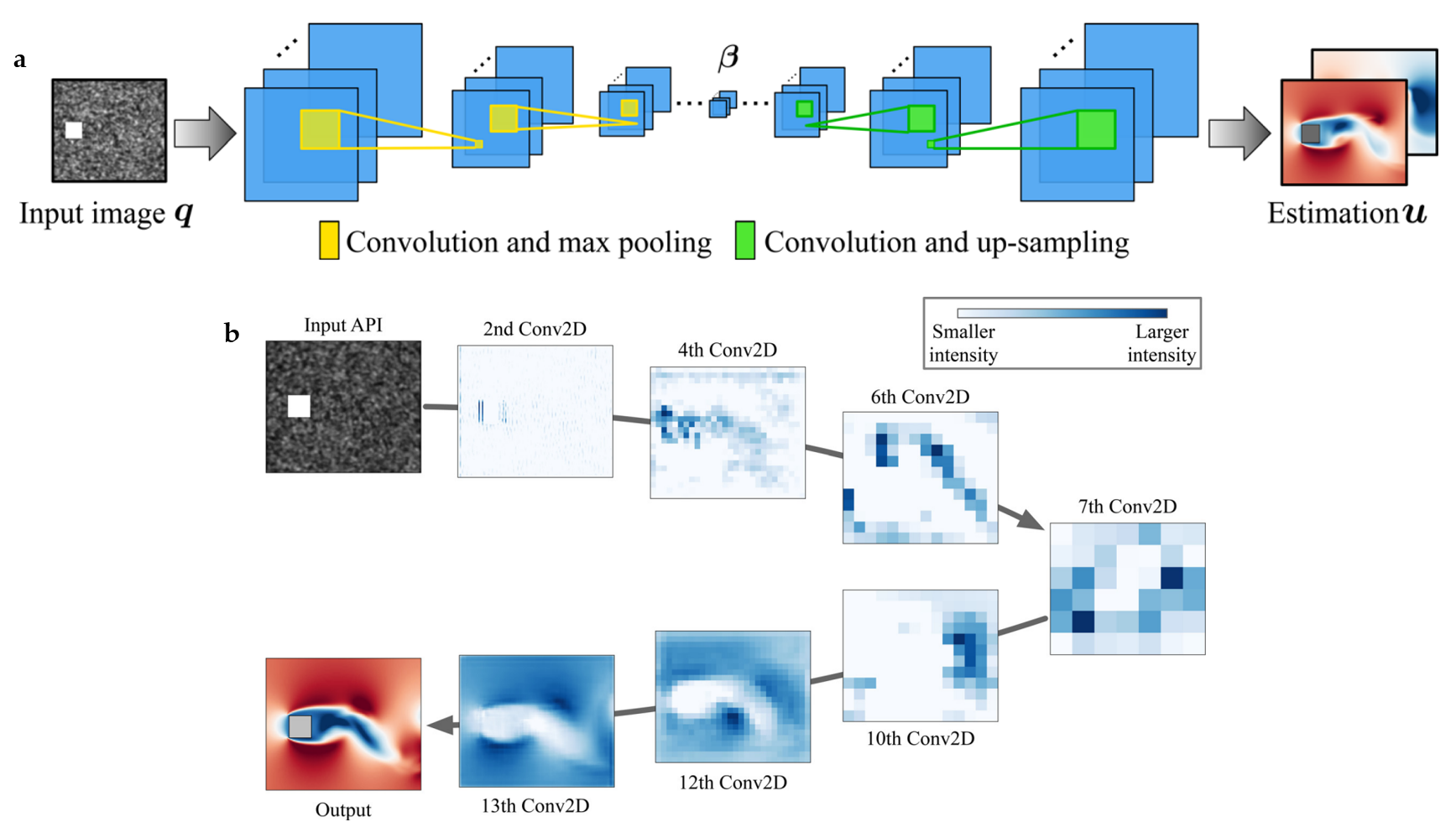}
   \caption{{\bf Process of PIV analysis based on ML.} {\bf a}, Representation of the autoencoder (AE) architecture based on convolutional neural networks (CNNs) to predict the flow field based on the input image containing the particle distribution. Yellow denotes convolution and max pooling, which reduce the size of the image, while green indicates convolution and up sampling, yielding an increase of the image size before the output. {\bf b}, Example of application of this method to the flow around a square cylinder. An artificial particle image (API) is used as input, and the output of each of the hidden layers is shown. The size of the image is progressively reduced until the 7th convolutional layer; beyond this point the size is increased successively until reaching the predicted velocity field in the output. Note that the first layers exhibit simpler and smaller features, which are hierarchically combined into larger, more complex elements when the output layer is approached. Figured adapted from Ref.~\cite{morimoto_et_al}, with permission from the publisher (American Institute of Physics).}
   \label{fig:morimoto}
\end{figure}

Another technique which can benefit from data-driven methods is particle-image velocimetry (PIV). In this technique, which enables obtaining instantaneous flow-field measurements, the flow is seeded with tracer particles which follow the dynamics of the stream. These particles are illuminated with a laser, and the resulting images are used to compute the instantaneous velocity vector~\cite{adrian_piv,scarano_piv,soria_piv}. This measurement technique can be enhanced in a number of ways through data-driven methods; for instance, M\'endez {\it et al.}~\cite{mendez_et_al} proposed a technique based on proper-orthogonal decomposition (POD)~\cite{lumley} to remove the noise and some of the inaccuracies in PIV measurements. 
Robust principal component analysis (RPCA) may also be used to remove outliers and fill in occlusions in PIV data, as demonstrated by Scherl et al.~\cite{Scherl2020prf}. 
Multi-layer perceptrons (MLPs) and CNNs were exploited by Rabault {\it et al.}~\cite{rabault_piv} to improve the post-processing method for PIV measurements and to provide super-resolution \cite{fukami2019super}. Furthermore, autoencoders (AEs) combined with CNNs were employed by Morimoto {\it et al.}~\cite{morimoto_et_al} to perform robust flow estimations, which can also be used to improve the quality of PIV measurements. A schematic representation of the approach adopted in this work is provided in Figure~\ref{fig:morimoto}.  
Other deep convolutional-neural-network architectures have also been explored for PIV enhancement~\cite{lee2017piv}. 

Another possible approach to enhance PIV measurements is to exploit the framework of physics-informed neural networks (PINNs)~\cite{raissi2019physics,raissi_et_al}, which is based on using the machinery for training neural networks (including the important advantages of the automatic differentiation) to solve partial differential equations (PDEs) through minimization of the PDE residual. This framework has exhibited great potential when solving certain PDEs, and enables embedding certain theoretical considerations (such as perturbation theory) to enhance its results~\cite{pinns_arzani}. In particular, PINNs have been shown to effectively improve measured flow fields with noise~\cite{pinns_exp}, and can also increase the resolution and quality of PIV measurements in turbulence~\cite{grauer}. The vortex-in-cell method~\cite{vic_ref}, which involves solving the vorticity-transport equation assuming inviscid flow and incompressibility, has been used to improve the temporal resolution of PIV measurements, including the resolution of the flow structures~\cite{vic_scarano}. Predictions of the coherent structures in turbulent flows have also been conducted via deep learning, and this methodology can improve the spatio-temporal representation of the coherent structures in PIV measurements. Note that a complete review of recent applications of machine learning to enhance PIV measurements can be found in the work by Discetti and Liu~\cite{discetti_rev}.


\section*{Experimental design and modeling enhanced by machine learning}

The next broad area of experimental fluid mechanics that stands to benefit from machine learning includes the design of experiments, developing reduced-order models, and uncertainty quantification.  Many of these approaches culminate in the digital twin~\cite{willcox_digital,niederer2021scaling,brunton2021data}, which is a digital model of the physical asset (i.e., the fluid flow experiment, or a closed-loop cyber-physical system, etc.).  These models are constantly updated with measurement data and may incorporate a hierarchy of physics-based and machine-learned models.  

When performing wind-tunnel experiments to study canonical flow cases there is an important aspect to take into account: how similar are the experimental realization and the canonical definition of the flow under study. The walls in wind-tunnel test sections introduce additional boundary layers which may affect the flow, for instance through blockage effects~\cite{blockage}. Furthermore, the corners in the test section introduce secondary flows of Prandtl's second kind, which may also impact the experimental results~\cite{modesti_duct}. Numerical simulations and empirical models can be used to design inserts in order to establish the desired flow evolution and pressure-gradient conditions~\cite{narges_fluids}; however, in many cases it is challenging to achieve those conditions, a fact that can significantly impact the quality of the measurements. In these situations it is possible to use data-driven methods to optimize the shape of the inserts so as to achieve a particular pressure-gradient distribution. In particular, Morita {\it et al.}~\cite{morita_et_al} proposed a framework based on Gaussian processes which enables conducting shape optimization on the inserts of a wind-tunnel setup. In their work, the framework not only converges within few iterations to the optimal insert design, but also decides the next case to explore in the parameter to space to achieve the fastest possible convergence. Note however that the Gaussian-process approach is a valid option only when a relatively low number of parameters are to be optimized. For higher-dimensional cases, gradient-based optimization~\cite{gradient_ref} can be a suitable choice. 

It is natural to synthesize experimental measurements into reduced-order models that may be used to predict, estimate, and control the behavior of both the fluid flow and the experimental apparatus.  
There is a wide variety of data-driven modeling algorithms that may be used to characterize a system, ranging from the linear dynamic-mode decomposition (DMD)~\cite{Rowley2009jfm,Schmid2010jfm,Kutz2016book,Rowley2017arfm}, higher-order DMD~\cite{hodmd} the interpretable sparse identification of nonlinear dynamics (SINDy)~\cite{Brunton2016pnas,Rudy2017sciadv}, genetic programming and symbolic regression~\cite{Bongard2007pnas,Schmidt2009science,cranmer2019learning,cranmer2020discovering}, and various neural-network-based modeling techniques~\cite{vlachas2018data,pathak2018model,cranmer2020lagrangian,chen2018neural}.  Sparse nonlinear modeling techniques have shown particular promise in identifying interpretable and generalizable models from experimental data~\cite{reinbold2021robust,Callaham2022scienceadvances,supekar2023learning} and designing closure models~\cite{zanna2020data,schmelzer2020discovery,beetham2020formulating,beetham2021sparse} that may be informed by experimental data in the future. 
Related parsimonious models were recently developed for an actuated D-shaped body in experiments, capturing all of the resonance and lock-on phenomena~\cite{herrmann2020modeling}. 

Another essential aspect of the experimental work in fluid mechanics is uncertainty quantification (UQ). Being able to add error bars to any results from an experiment is critical in order to fully interpret and understand the reported results. In the traditional approach to UQ~\cite{bailey_et_al} the method of propagation of errors is used, where the uncertainty of a particular measured quantity is essentially the addition of the uncertainties from the individual quantities used to calculate the first one. More recent studies have proposed comprehensive UQ frameworks that can capture more nuanced interactions among the measured quantities~\cite{saleh_exp}, providing a more robust assessment of the uncertainty. Using the example of hot-wire anemometry, Ref.~\cite{saleh_exp} employs Monte-Carlo (MC) sampling~\cite{sobol}, polynomial-chaos expansions (PCE)~\cite{xiu_et_al} and linear-perturbation methods (all providing similar results) to take into account the correlations among variables when assessing the uncertainty. They found that, due to error cancellation, these methods provide lower uncertainties than those reported using propagation of errors. Furthermore, Sobol indices~\cite{sobol} enable assessing the sensitivity of a particular measurement with respect to the various intermediate quantities measured to compute the first one. In the case of hot wires, the wire voltage is the quantity with highest impact on the results. When it comes to oil-film interferometry (OFI)~\cite{tanner_blows}, which is a technique used for direct measurement of the wall-shear stress $\tau_w$, the quantity with the largest effect on the measurement of $\tau_w$ is in fact the calculation of the so-called fringe velocity. This is a complicated process~\cite{nagib_ofi} requiring extensive image analysis, thus computer-vision methods such as convolutional neural networks (CNNs)~\cite{lecun} may help to improve the procedure. Note that advanced UQ methods are becoming the standard also in the context of computational fluid dynamics (CFD)~\cite{saleh_cfd}.

\begin{figure}
\centering 
\includegraphics[width=0.995\textwidth]{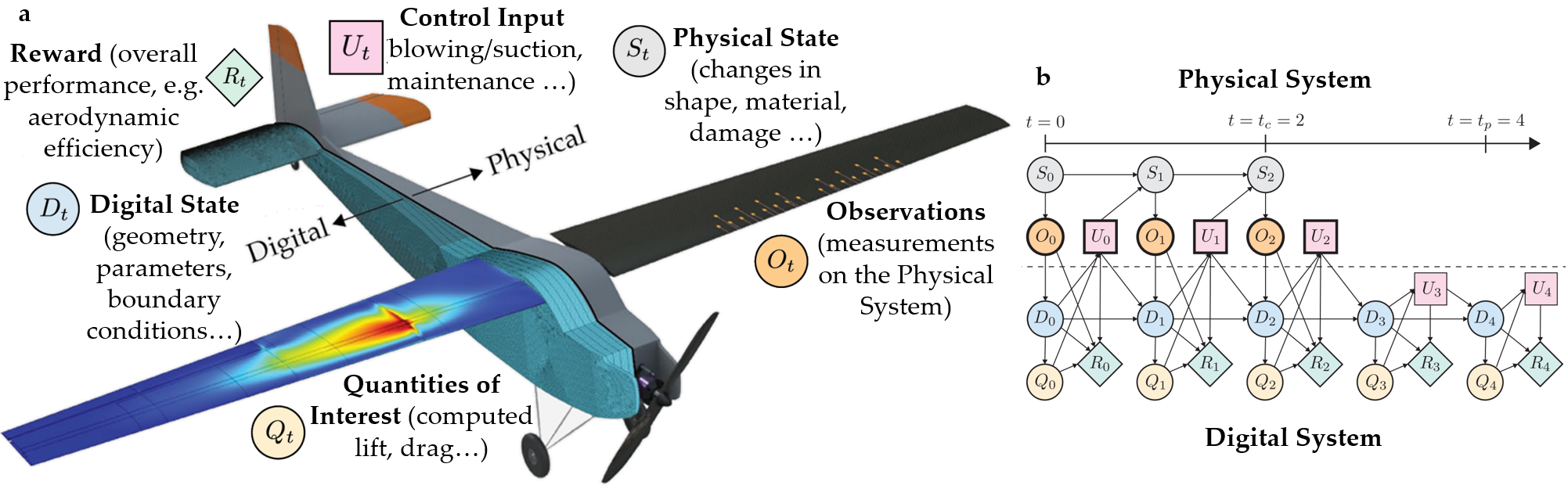}
   \caption{{\bf Schematic representation of a digital twin.} {\bf a}, Digital twin illustrated in a simple aircraft, where the Observations, the Physical State and the Control Input are quantities belonging to the Physical System, whereas the Digital State, the Quantities of Interest and the Reward are associated with the Digital System. {\bf b}, Example of application of this framework with a dynamic-decision network, where the quantities above the dashed line correspond to the Physical System and the ones below to the Digital System. Here the complete Physical State (which may not be completely observable) defines the Observations in the Physical System (e.g. wall-shear stress or wall pressure). The observations enable computing the Digital State through numerical models, which in turn can be used to calculate the quantities of interest and define the overall reward. The framework also allows eventually evolving the solution in time without experimental measurements. Figure adapted from Ref.~\cite{willcox_digital}, with permission from the publisher (Springer Nature).}
   \label{fig:kapteyn}
\end{figure}

Reduced-order models, informed and updated by data with quantifiable uncertainty, culminate in the so-called digital twin~\cite{willcox_digital,niederer2021scaling,brunton2021data}. 
Digital twins are essentially a virtual representation of a physical asset, typically the result of a hierarchy of models of different fidelity, and constantly being updated by various data streams from the real world.  
The digital twin comprises physical, machine learning, and hybrid models.  
Such a digital representation of a complex physical system may enable major advances in optimization, iterative design, and control.  
A detailed schematic of the digital-twin methodology applied to a simple aircraft, including all the variables and the dynamic-decision network, is provided in Figure~\ref{fig:kapteyn}. 
Because digital twins balance models and data, it is important to effectively sample the system of interest, requiring modern active learning~\cite{fan2019robotic} and design of experiments~\cite{qr_pivoting}. 
Further, a hierarchy of classical physics-based models and modern data-driven reduced-order models will likely be essential for characterizing a digital twin.

\section*{Possibilities within data-driven flow estimation and control}

Finally, there are tremendous opportunities for machine learning to advance the challenging problems of sensing/estimation and control of fluid flows.  An improved ability to model and control fluid flows may be enabler of advanced technology, and experimental demonstrations are necessary before wide-scale deployment.  
However, the field of fluid dynamics poses several challenges for control~\cite{Brunton2015amr,Brunton2020arfm}.  
The dynamics are typically high-dimensional, nonlinear, and multiscale in both space and time.  
Further, relevant timescales are often quite fast for modern applications, requiring fast control decisions to reduce latency.  
There are also often strong instabilities and time delays between sensing and actuation.  
Thus, there are opportunities for better sensor and actuator placement, improved model-based estimation from limited and noisy sensors and enhanced closed-loop flow control, all enabled by machine learning.

Data-driven methods are enabling a number of possibilities within sensing in experiments. This is an essential ingredient of reactive control, where being able to accurately assess the state of the flow based on sparse measurements significantly improves the possibility of performing effective flow control. Some traditional approaches to sense the flow based on information at the wall include linear stochastic estimation (LSE)~\cite{encinar_lse,hasegawa_lse} and extended proper-orthogonal decomposition (EPOD)~\cite{boree_epod} (which are formally equivalent). Note that turbulent flows are inherently nonlinear, and although some of the energy-transfer phenomena are linear (namely, superposition~\cite{modulation1}), the modulation is a nonlinear phenomenon~\cite{modulation3} which cannot be accurately reproduced by linear methods. More recently, non-linear transfer functions have been used to make such predictions with improved results, and the potential of computer-vision-based methods such as convolutional neural networks (CNNs) has been assessed in turbulent channels~\cite{guastoni2}. CNNs are able to exploit the spatial information in the input for obtaining improved predictions, and given the pronounced signature of the turbulence structures at the wall, these neural networks can successfully predict the flow up over a wide range of wall-normal locations. Using the wall-shear-stress components and the wall pressure as inputs, Guastoni {\it et al.}~\cite{guastoni2} predicted the turbulence fluctuations at $y^+=15$ (where $y^+$ is the wall-normal coordinate in inner scaling) with less than $1\%$ error using CNNs, and with less than $30\%$ at $y^+=100$. Some variants of this approach have been proposed in the literature, including combining CNNs with POD~\cite{guemes_fcn_pod}, or using deep CNNs for predictions based on the heat flux~\cite{kim_lee}. The limited spatial resolution encountered in experimental settings, which may complicate the usage of CNNs, was circumvented by G\"uemes {\it et al.}~\cite{guemes_gans} by using generative adversarial networks (GANs)~\cite{goodfellow2020generative}. In their proposed architecture, super resolution is used to produce highly-resolved representations of the data at the wall, which are in turn used to produce accurate predictions of the turbulent fluctuations above the wall. These sensing approaches, when conducted in real time, can guide the wall actuation to target certain scales in the flow, with the aim of achieving drag reduction~\cite{baars_sensing}. Note that, although very promising results have been obtained from numerical data, sensing in actual experimental applications is associated with additional challenges, {\it e.g.} due to the filtering effect of the measurement techniques (beyond the reduced resolution), which may lead to a nontrivial implementation of the methods developed from numerical data.

Another relevant area within sensing in experiments is the enhancement of measured flow quantities. For instance, numerical data can be used to train deep neural networks to correlate the flow with the polymeric stresses in non-Newtonian flows~\cite{AriNonNewtonian}; then, these deep-learning models can be used to obtain, in an experimental setup, those polymeric stresses based on flow quantities (which are relatively straightforward to measure). In connection with this, recent work based on graph neural networks has enabled learning the rheologic parameters of complex flows through a reduced number of experiments~\cite{graph_rheology}. These approaches may enable much deeper physical insight into the dynamics of non-Newtonian flows, with very important implications in a wide range of industrial applications. On the other hand, it is essential to establish methodologies for optimal sensor placement, particularly in complex geometries. Given the limited resolution and amount of sensors available in experimental setups, one can devise strategies for optimal sensing using data-driven methods. A very effective way to optimize sensor locations in high-dimensional systems is the so-called QR pivoting~\cite{qr_pivoting}, which involves reduced-order-model (ROM) methods such as singular-value decomposition (SVD)~\cite{Brunton2022book}. These techniques rely on classical tools from linear algebra, and provide excellent results in a wide range of applications. However, turbulent flows are inherently nonlinear, which means that it is convenient to employ methods involving nonlinearity for sensor placement. In this direction, the tools for explainability of neural networks can provide valuable insight into optimal sensor placement. Explainability implies being able to evaluate the parts of the input data that contribute the most to the prediction of the output~\cite{vinuesa_interp}, and one popular framework to evaluate this is based on the so-called SHapley Additive exPlanations (SHAP) values. These values, which rank the features of the input by their prediction importance, can be calculated by means of the kernel SHAP method~\cite{shap}. If a deep neural network is trained to predict flow quantities based on features measured at the wall, the SHAP framework can provide a map of importance of all the wall points, effectively leading to a framework to decide the optimal location of point sensors while leveraging the nonlinearities of deep neural networks.

Flow control, both in its active and passive forms, can also benefit from the high-dimensional optimization capabilities of data-driven methods. On the one hand, it is possible to use machine learning (and exploit the capabilities of transfer learning~\cite{guastoni2}) to fine tune the correlations used to define the geometry of the wall roughness, in order to establish effective passive-control methods~\cite{shervin_transfer}, with application to experimental setups. On the other hand, several approaches have been used to optimize active flow control. For instance, Mahfoze {\it et al.}~\cite{gp_control} employed Bayesian optimization to identify the best combination of control-region length and blowing amplitude to maximize the energy savings, in an approach which also included intermittent control regions. Furthermore, these authors also took into account the data by Kornilov~and~Boiko~\cite{KornilovBoiko} to formulate a more realistic estimate of the power consumption by blowing. Another data-driven control approach which has been proved successful in controlling external flows in experimental settings is genetic programming~\cite{noack_control,minelli_et_al}. This technique enables developing novel control strategies with a certain degree of interpretability in the resulting control law. When it comes to turbulence control, there is indeed an interesting question regarding which scales need to be attenuated to achieve drag reduction. While the traditional opposition control~\cite{choi_et_al} focuses on the near-wall fluctuations, recent work by Marusic {\it et al.}~\cite{marusic_et_al} indicates that at higher Reynolds numbers larger scales need to be controlled. 

\begin{figure}
\centering 
\includegraphics[width=0.9\textwidth]{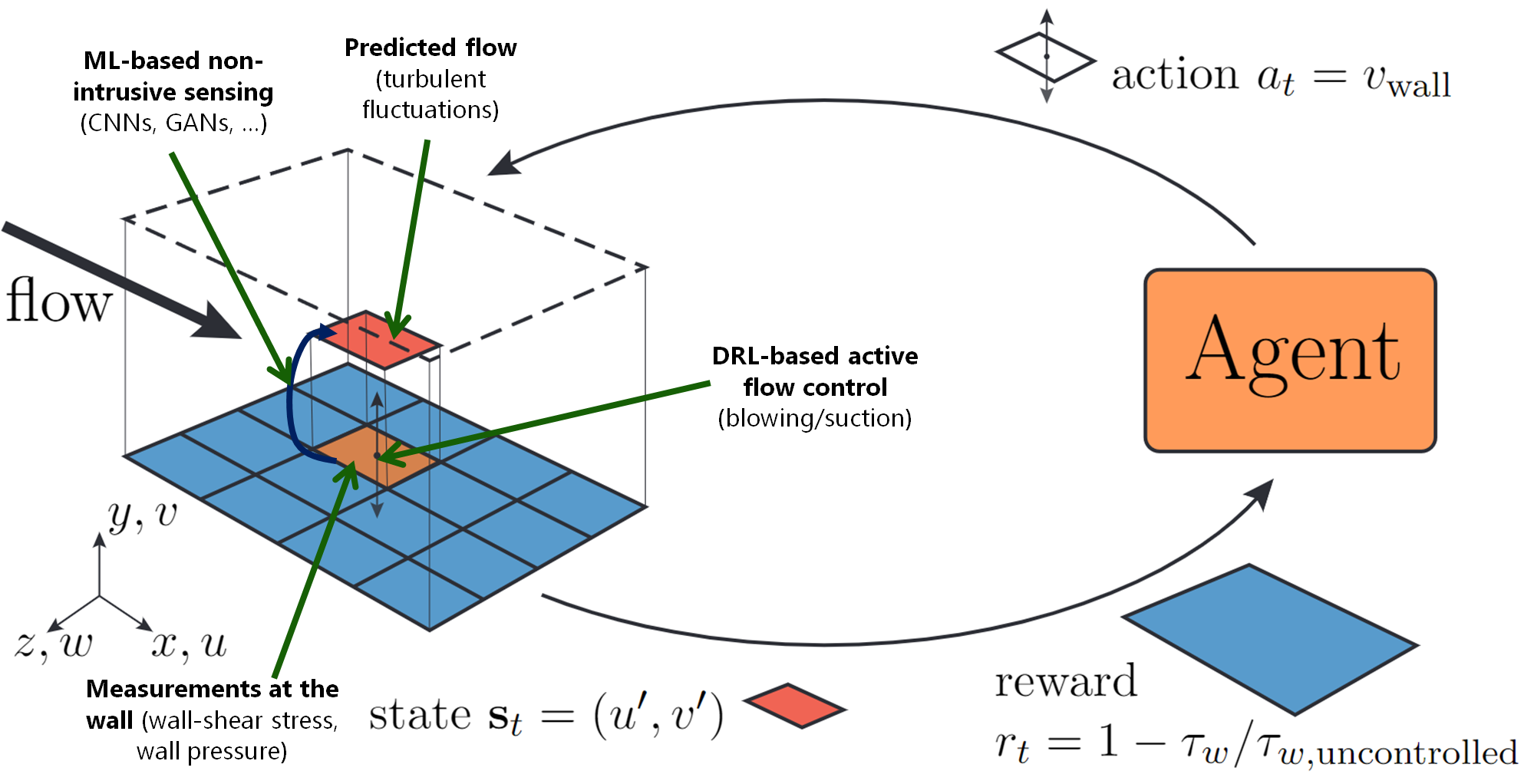}
   \caption{{\bf Schematic representation of the process for sensing and control.} In this figure we show a turbulent open channel, where the blue surface represents the wall and at the upper dashed plane a symmetry condition is imposed. Based on measurements at the wall (orange panel), which are non intrusive, it is possible to use computer-vision methods (such as convolutional neural networks, CNNs, or generative adversarial networks, GANs) to predict the instantaneous velocity fluctuations above the wall (red panel). Having this information, a deep-reinforcement-learning (DRL)-based algorithm determines the optimal control policy to minimize the wall-shear stress. The arrows represent active flow control by means of blowing and suction. Figure adapted from Ref.~\cite{GuastoniDRL}.}
   \label{fig:guastoni}
\end{figure}

Reinforcement learning (RL) is a promising approach at the intersection of machine learning and control~\cite{Sutton1998book,recht2019tour,mnih2015human}, where it has been widely applied to solve several challenges in gaming and general artificial intelligence~
\cite{silver2018general,reddy2018shared,vinyals2019grandmaster}.  
Several RL approaches have been explored in the fluid-mechanics literature~\cite{verma2018efficient,novati2019controlled,rabault2019artificial,noack_drl,fan2020reinforcement,novati2021automating,gunnarson2021learning,vinuesa_fluids,bae_koumoutsakos,GuastoniDRL,HasegawaDRL,vignon}, to varying degrees of success.  
Deep reinforcement learning (DRL), where deep neural networks are used for some aspect of the RL process, has been particularly powerful, given sufficient training resources.  
In this sense, DRL is an approach that can help to establish novel and effective control strategies~\cite{vignon}, discerning the ranges of scales to be targeted. More recently, DRL has been shown to provide very promising drag-reduction rates in turbulent channel flow~\cite{GuastoniDRL,HasegawaDRL}. In particular, Guastoni {\it et al.}~\cite{GuastoniDRL} reported a larger drag reduction ($30\%$) using DRL than that obtained via opposition control ($20\%$). A very promising (and challenging) experimental application can be obtained by combining non-intrusive sensing~\cite{guastoni2,guemes_gans} and DRL-based flow control~\cite{GuastoniDRL,HasegawaDRL}, as illustrated in Figure~\ref{fig:guastoni}.

\section*{Conclusions and outlook}
In this perspective review, we have summarized a number of promising avenues of experimental fluid mechanics that are being transformed by techniques in machine learning.  
In particular, we have explored efforts to 1) augment the fidelity of existing measurement techniques, 2) improve experimental design and the surrogate modeling of experiments and 3) enable real-time estimation and control.  

Augmenting measurement fidelity is a natural application of machine learning, where similar advances have been made in image processing.  For example, super-resolution, outlier detection, and data imputation are all existing ML techniques that have been shown to have immediate benefit in improving fluid velocity fields from PIV.  
The addition of known physics, through regularizing terms and constraints has further improved the ability to infer physically-consistent flow fields from limited measurements.  
This area of research is likely to continue rapid development, as it improves the measurement capabilities without more costly experimental equipment, by simply leveraging the wealth of data available from other high-fidelity measurements and computations.  
The synthesis of fluid databases into larger \emph{foundation models} for fluid mechanics will be a crucible for these efforts.  

The second area of high-priority research is in improving experimental design and the construction of digital-twin surrogate models for experiments.  
Characterizing a fluid-flow experiment across a range of relevant parameters (e.g., Reynolds number, wing geometry, etc.), is often important for inverse design, although it may be prohibitively expensive to sweep through all parameters.  
Instead, emerging techniques in active learning are providing a principled approach to sample these parametric spaces efficiently while providing error bounds and uncertainty estimates on the resulting models.  
The resulting surrogate models may then be used flexibly in iterative design optimization and downstream estimation and control tasks.  
The development of surrogate models is another area that is being dramatically improved with machine learning techniques.  
All of these approaches culminate in the \emph{digital twin}, which is a hierarchy of models (based on physics, machine learning, or hybrid) that are pinned to measurement data and come with uncertainty quantification.  
Digital-twin pipelines are still being developed and there are many open questions.  However, the tremendous industrial benefit of model-based engineering is likely to continue driving these efforts for decades.  

Finally, we have explored the potential of machine learning to improve real-time estimation and control in fluid experiments.  
Estimation and control are essential tasks, both to characterize and regulate an experiment as well as an objective of the flow experiment itself (e.g., to minimize drag or maximize lift).  
Flow control is often limited by computational latency, and machine-learning solutions are notoriously fast, balancing accuracy and efficiency.  
Although major advances have been made in the field of flow control, several challenges remain, including future control plants with many sensors and actuators, resulting in a high-dimensional search space.  
Further, strong nonlinearities and time delays make control especially challenging, although advances in reinforcement learning and ML-enabled model predictive control (MPC) are promising.  
It is important to note that when possible, simpler optimization and control strategies may also be advantageous, for example in the use of simplex optimization to improve the power efficiency of a lab-scale cross-flow turbine by over 50\%~\cite{Strom2017natenergy}. 

To summarize the various possible applications of ML to complement and enhance certain aspects of experimental fluid mechanics, in Table~\ref{tab:methods} we present a selection of potentially relevant ML methods. This table is meant to inform an experimentalist on the type of approach that might be more convenient depending on their intended problem, so that it becomes easier to select the most suitable ML algorithm from the vast catalog of possibilities.

\begin{table}
\centering 
\includegraphics[width=0.995\textwidth]{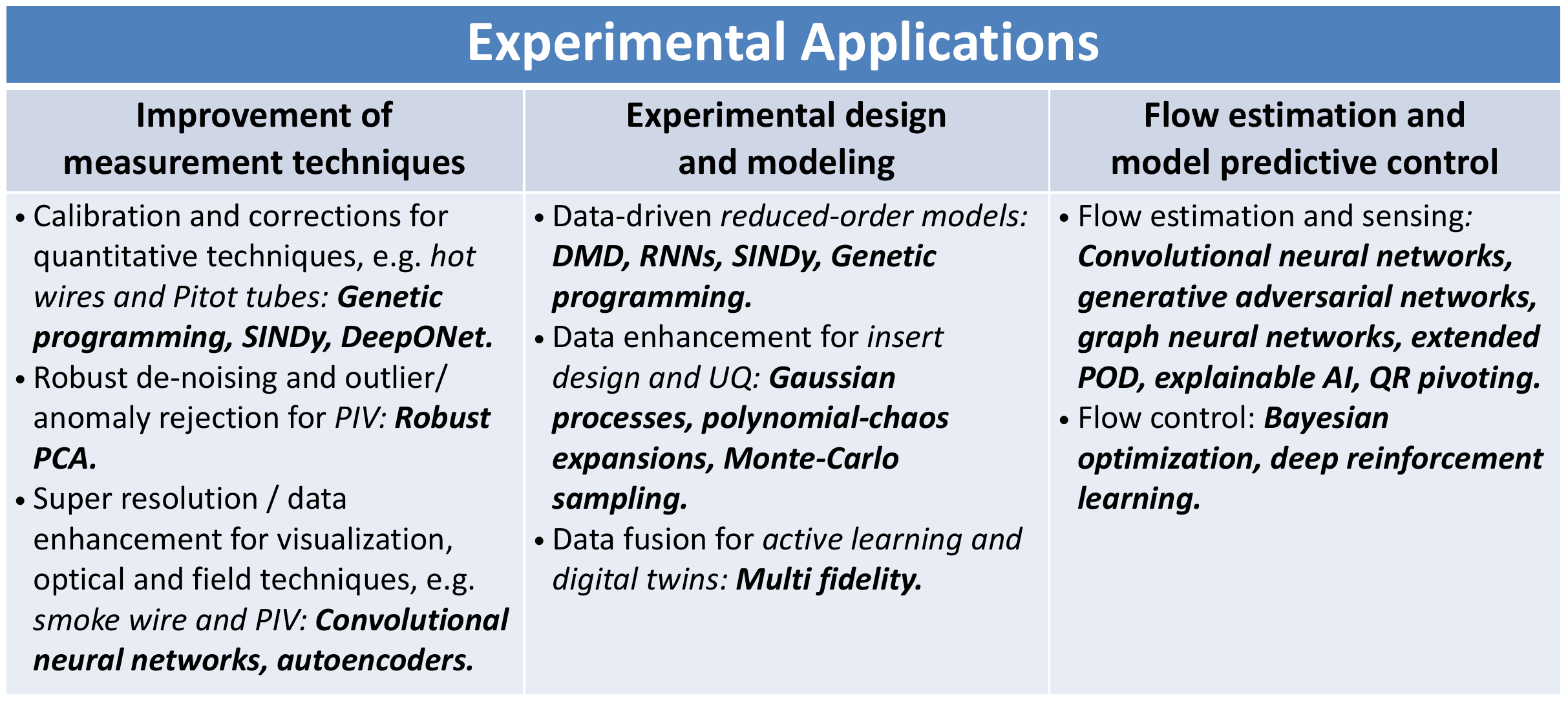}
  \caption{{\bf Summary of sample ML methods to use for various applications within experimental fluid mechanics.}  We show in italics experimental techniques or tasks, and in bold face relevant ML methods.}
   \label{tab:methods} 
\end{table}

Beyond these high-profile uses of machine learning in experimental fluid mechanics, there are several other emerging opportunities and applications.  Cloud connectivity is making it possible to have remotely-controlled, collaborative fluid-flow facilities, perhaps providing a new generation of shared experimental resources.  

For the foreseeable future, experiments will remain the gold-standard in fluid mechanics for capturing true multiscale, multiphysics effects for flows with complex geometries.   
Increasing measurement fidelity and machine-learning algorithms will likely make these experiments even more relevant, with the possibility to more flexibly transfer results from one configuration to another, which is a standing open problem.  
With fluids at the center of several trillion-dollar industries, from health and defense to energy and transportation, it is expected that these technologies will continue to mature rapidly, driven by considerable industry investment.  
The aerospace industry and biomedical applications have been particularly strong early adaptors, and it is believed that digital twins with fluid flows will continue to develop in these fields. Furthermore, the great relevance of fluid mechanics for the current climate emergency further justifies the adoption of these novel approaches to accelerate related scientific discovery.

\section*{Acknowledgements}
The authors would like to gratefully acknowledge valuable discussions with Bernd Noack early in the development of this perspective. 
RV acknowledges financial support from ERC grant no. `2021-CoG-101043998, DEEPCONTROL'. Views and opinions expressed are however those of the author(s) only and do not necessarily reflect those of the European Union or the European Research Council. Neither the European Union nor the granting authority can be held responsible for them. 
SLB acknowledges support from the National Science Foundation AI Institute in Dynamic Systems
(grant number 2112085).  BJM is grateful for the support of the U.S. ONR through a Vannevar Bush Faculty Fellowship,  N00014-17-1-3022.

 \begin{spacing}{.88}
 \setlength{\bibsep}{2.pt}
\bibliographystyle{abbrvnat}
\bibliography{ai4exp_bib}
 \end{spacing}
\end{document}